# Numerical recovering a density by BC-method

In this paper we develop numerical algorithm for solving inverse problem for the wave equation using Boundary Control method. The results of numerical experiments are represented.

Leonid Pestov Ugra Research Institute of Information Technologies Khanty-Mansiysk, 628011, Russia

Victoria Bolgova and Oksana Kazarina Ugra State University Khanty-Mansiysk, 628000, Russia

## 1 Introduction

The Boundary Control method is one of the most natural method for solving multidimensional inverse dynamical problems. The method was proposed by Belishev in 1986 (see [B], [B1] and publications cited there). Numerical testing results are represented in [BG].

The principle question in the BC-method is the controllability. For scalar equations like wave equation the approximate controllability is valid. It means any function $\varphi$ in bounded domain $\Omega$ (say from $L^2(\Omega)$) may be arbitrary close approximated by a wave $u^f(.,T)$, induced by some boundary source $f$ (control). For arbitrary $\varphi$ the equation $u^f(.,T)=\varphi$ w.r.t. $f$ cannot be solved when coefficients of the (wave) equation are unknown. The remarkable fact is that for harmonic $\varphi$ one can find a control which provides the equality $u^f(.,T)=\varphi$ with arbitrary accuracy using only data of the inverse problem. But in practice we have only finite number of controls. Can one provide proper accuracy at that and how many controls is takes? In this paper we try to answer these questions numerically when reconstructing a density in the unit disk in $R^2$. We take $T$ more than optical radius of $\Omega$. It makes our problem easier.

We prove also approximate $H^1$-controllability. We use this to develop a numerical algorithm for solving inverse problem for the wave equation with unknown density. We also demonstrate the results of numerical experiments.

Let $\Omega$ be a bounded one-connected domain in $R^n$ ($n=2,3$) with a smooth boundary $\Gamma$. Denote by $u^f$ the solution (*wave*) to the initial boundary problem for the wave equation

$$\rho u_{tt}-\Delta u=0 \;\; in\, \Omega\times(0,2T) \tag{1}$$

$$u|_{t=0}=u_t|_{t=0}=0, \tag{2}$$

$$u_n|_{\Gamma\times 0,T]}=f\in F^T=\{f\in L^2(\Gamma\times 0,2T]),\ f|_{t>T}=0\}, \tag{3}$$

where $\rho(x)>0$ is a smooth density, $u_n$ is the normal derivative (*control*). Here $L^2(\Gamma\times 0,2T])$ is real (as all Hilbert spaces in this paper). The map $R^{2T}:F^T\to L^2(\Gamma\times 0,2T])$ defined by

$$R^{2T}f=u^f|_{\Gamma\times 0,2T]}$$

is bounded [L] and called *response operator*. The wave $u^f$ is classical when $f\in \mathrm{M}$,

$$\mathrm{M}=\{f\in C^\infty(\Gamma\times 0,2T])\,|\,f|_{t=0}=f_t|_{t=0}=0\}.$$

Consider inverse problem: to reconstruct $\rho$ from $R^{2T}$ under assumption

$$T > T^{*} = sup_{x\in\Omega}\ dist\,(x,\Gamma), \tag{4}$$

where the distance is taking with respect to the metric $\sqrt{\rho(x)}|dx|$. This assumption provides that the waves induced by various controls fill up the closure of $\Omega$ at the final moment $T$. Note, that if $0 < T < T^{*}$ then data $(f, R^{2T} f),\ f \in F^{T}$ of the the inverse problem do not include information about the density on the set $\overline{\Omega}\setminus\Omega^{T}$, where

$$\Omega^{T} = \{x \in \overline{\Omega} \mid T^{*} \le T\}$$

is the set filled by waves up to the moment $T$. In our case $\Omega^{T} = \overline{\Omega}.$ In what follows the fixed final moment $T > T^{*}$ will not be mentioned at all notations.

Our approach to this problem is based on the BC-method and close to approach of [P].

## 2 Bilinear forms

Introduce two symmetric bilinear forms

$$[f,g] = \int_{\Omega}\rho u^{f}(.,T)u^{g}(.,T)dx, \tag{5}$$

$$[f,g]_{p} = \int_{\Omega}(\nabla u^{f}(.,T),\nabla u^{g}(.,T))dx.$$

The following relations between these forms and response operator are the base of BC-method. We obtain these relations for the convenience of riders (for more details see [B]). We use the notations

$$u_{\pm}(.,t) = (u(.,t) \pm u(.,2T-t))/2,$$

$$(If)(.,t) = \int_{0}^{t} f(.,s)ds,\ \ t \in 0,2T],$$

$d\Sigma$ is the volume form of $\Gamma\times 0,T]$.

**Proposition 1** *For any controls* $f, g \in \mathrm{M}$ *the equalities*

$$[f,g] = \int_{\Gamma\times 0,T]}[(Rg)_{+} If - g_{+} IRf]d\Sigma, \tag{6}$$

$$[f,g]_{p} = \int_{\Gamma\times 0,T]}[f\frac{\partial}{\partial t}(Rg)_{+} + g_{+}\frac{\partial}{\partial t}Rf]d\Sigma, \tag{7}$$

are valid.

**Proof.** For any solution $v$ to the wave equation the equality

$$\rho(vu_{t}^{f} - u^{f}v_{t})_{t} = div(v\nabla u^{f} - u^{f}\nabla v) \tag{8}$$

holds. Substituting $u_{+}^{g}$ for $v$ and integrating over cylinder $\Omega\times 0,T]$, we get

$$\int_{\Omega}\rho u^{g}(.,T)u_{t}^{f}(.,T)dx = \int_{\Gamma\times 0,T]}[u_{+}^{g} f - u^{f}(u_{+}^{g})_{n}]d\Sigma.$$

Taking into account that $u^{f_t} = u_{t}^{f}$ and $u^{If} = Iu^{f}$, we get (6).

Consider (potential) bilinear form $[f,g]_{p}$. For any solution $v$ to the wave equation the equality

$$\rho(v_{t}u_{t}^{f})_{t} + (\nabla v,\nabla u^{f})_{t} = div(v_{t}\nabla u^{f} + u_{t}^{f}\nabla v).$$

holds. Substituting $u_{+}^{g}$ for $v$ and integrating over $[\Omega\times 0,T]$, we get

$$(\nabla u^g, \nabla u^f)(x,T)dx = \int_{\Gamma\times 0,T]}[(u_+^g)_t f + u_t^f (u_+^g)_n]d\Sigma.$$

**Remark 2** *Analogously to (7) one can get representation of the kinetic form*

$$[f,g]_k \overset{def}{=} \int_\Omega \rho(x) u_t^f(x,T) u_t^g(x,T)dx = \int_{\Gamma\times 0,T]}[f\frac{\partial}{\partial t}(Rg)_- + g_- \frac{\partial}{\partial t}Rf]d\Sigma. \tag{9}$$

**Remark 3** *It may be shown (6),(7),(9) do not depend on values of controls on the set* $\Gamma\times T,2T]$.

## 3 Boundary control and density reconstruction

Consider the boundary control problem

$$u^f(.,T) = \varphi \in L^2(\Omega),\ f = ? \tag{10}$$

For any $T>0$ the linear variety $\mathrm{H} = \{u^f(.,T) \mid f \in \mathrm{M}\}$ is dense in $L^2(\Omega)$ [B]. For sufficiently large $T > T^*$ and under some geometrical assumption the equation is solvable in $F$ (not uniquely) [BLR].

We consider the case when $\varphi \in H^1(\Omega)$, where $H^1(\Omega)$ is the real Hilbert space with the norm

$$\|u\|_1 = (\int_\Omega \rho u^2 dx + \int_\Omega |\nabla u|^2\ dx)^{1/2}.$$

Show that the set $\mathrm{H}$ is dense in $H^1(\Omega)$.

**Theorem 4** *The orthogonal complement to* $\mathrm{H}$ *in* $H^1(\Omega)$ *is* $\{0\}$.

**Proof.** Let $v$ be the solution to the initial boundary value problem

$$\rho v_{tt} - \Delta v = 0, in\ \Omega\times(0,T),$$
$$v_n|_{\Gamma\times 0,T]} = 0,$$
$$v|_{t=T} = 0,\ v_t|_{t=T} = \varphi,$$

where $\varphi \in \mathrm{H}^\perp$. Since $\varphi \in H^1(\Omega)$ we have

$$v \in C([0,T]; H^2(\Omega)),\ v_t \in C([0,T]; H^1(\Omega)),\ v_{tt} \in C([0,T]; L^2(\Omega))$$

[LM]. Denote by $u$ the wave $u = u^f - u_{tt}^f, f \in \mathrm{M}$ . By standard way

$$\int_\Omega \rho(uv_t - vu_t)(.,T)dx = \int_{\Gamma\times 0,T]}(uv_n - vu_n)d\Sigma \Rightarrow$$
$$\int_\Omega \rho u^f(.,T)\varphi dx - \int_\Omega \Delta u^f(.,T)\varphi dx = \int_{\Gamma\times 0,T]} v(f_{tt} - f)d\Sigma \Rightarrow$$
$$\int_\Omega \rho u^f(.,T)\varphi dx + \int_\Omega (\nabla u^f(.,T), \nabla\varphi)dx = \int_\Gamma \varphi f(.,T)d\Sigma + \int_{\Gamma\times 0,T]} v(f_{tt} - f)d\Sigma$$
$$= \int_{\Gamma\times 0,T]}(v_t f_t + vf)d\Sigma.$$

Thus we have

$$(u^f,\varphi)_{H^1(\Omega)} = \int_{\Gamma\times 0,T]}(v_t f_t + vf)d\Sigma.$$

Therefore $\varphi \in \mathrm{H}^{\perp}$ implies $v|_{\Gamma\times 0,T]} = 0$. The even extension of $v$ w.r.t. $t = T$

$$\widetilde{v}(.,t) = \begin{cases} v(.,t), \ t \in 0,T] \\ -v(.,2T-t), \ t \in T,2T] \end{cases}$$

is the function from $C([0,T]; H^2(\Omega))$ which satisfies

$$\rho v_{tt} - \Delta v = 0, in\ \Omega \times (0,2T),$$
$$v|_{\Gamma\times 0,2T]} = 0, \ \ v_n|_{\Gamma\times 0,2T]} = 0.$$

As in [B] using Tataru's theorem [T] this implies $v = 0$ in the domain bounded by characteristics $t = dist(x,\Gamma)$ and $t = 2T - dist(x,\Gamma)$. Thus $\varphi = 0$.

Let $\varphi$ be an arbitrary smooth harmonic function in $\Omega \cup \Gamma$. Consider the functional $\Phi : H^1(\Gamma \times 0,T]) \to R$:

$$\Phi(f) \overset{def}{=} \int_\Omega |\nabla u^f(.,T) - \nabla\varphi|^2\, dx$$
$$= \int_\Omega |\nabla u^f|^2 (.,T)dx - 2\int_\Omega (\nabla u^f(.,T), \nabla\varphi)dx + \int_\Omega |\nabla\varphi|^2\, dx$$
$$= [f,f]_p - 2\int_\Gamma (Rf)(x,T)\varphi_n(x)d\Gamma + \int_\Gamma \varphi\varphi_n d\Gamma.$$

Note, that $\Phi$ is completely determined by response operator. The following propositions is the base of our approach to reconstruct $\rho$.

**Proposition 5** *For any smooth harmonic function* $\varphi$ *and* $\varepsilon > 0$ *there is a control* $f \in \mathrm{M}$ *such that*

$$\Phi(f) + \|Rf(.,T) - \varphi\|^2_{L^2(\Gamma)} \le \varepsilon. \qquad (11)$$

Inequality (11) implies

$$\|u^f(x,T) - \varphi\|^2_{H^1(\Omega)} \le C\varepsilon,$$

with some constant $C$, does not depending on $f$ and $\varphi$.

**Proof.** The first statement follows from the density $\mathrm{H}$ in $H^1(\Omega)$ and the definition of $\Phi$. The second one follows from Friederischs' inequality.

Thus one can control the closeness $u^f(.,T)$ to $\varphi$ in $H^1$ from the boundary.

The reconstruction of the density may be fulfilled by the following scheme.

1. For any smooth harmonic functions $\varphi$ one can find the control $f_\varphi \in \mathrm{M}$ such that $\Phi_1(f_\varphi) + \|Rf_\varphi(.,T) - \varphi\|_{L^2(\Gamma)} \le \varepsilon$ and therefore $\|u^{f_\varphi}(x,T) - \varphi\|_{H^1(\Omega)} \le C\varepsilon$ with arbitrarily small $\varepsilon > 0$.

2. Substituting $f_{\varphi_1}$ for $u^f$ and $f_{\varphi_2}$ for $u^g$ in (5) we get approximate equality

$$\int_\Omega \rho(x)\varphi_1(x)\varphi_2(x)dx \approx [f_{\varphi_1}, f_{\varphi_2}], \qquad (12)$$

where $\varphi_1, \varphi_2$ are arbitrary smooth harmonic functions. Since the linear span of all products $\varphi_1\varphi_2$ is dense in $L^2(\Omega)$ then one can use (12) to find $\rho$. When numerical solving the inverse problem we use also the a priori limitations for $\rho$. Emphasize that both procedures are linear.

## 4 Discrete inverse problem

Project the forward problem (1-3) onto a finite dimensional space that spans standard continuous piecewise basic functions $\psi_n(x), n=1,\ldots, N$ of the Finite Element Method ($N$ is the number of nodes). Then (1-3) reduces to the Cauchy problem for a linear system of ordinary differential equations.

The projection $u_N^f$ of the wave $u^f$ is expended into the finite sum

$$u_N^f(x,t)=\sum_{n=1}^{N} U_n^f(t)\psi_n(x),\ \psi_n(x_m)=\delta_{nm}, n,m=1,\ldots, N,$$

where $x_n$ are nodes. Vector-function $U^f$ is the solution of the Cauchy problem for ordinary differential equations with constant coefficients :

$$MU_{tt}+KU=G^f,\ G_i^f(t)=\int_\Gamma \psi_i f(.,t)d\Gamma \quad (13)$$
$$U(0)=U_t(0)=0.$$

where

$$M_{ij}=\int_\Omega \rho\psi_i\psi_j dx,\ K_{ij}=\int_\Omega (\nabla\psi_i,\nabla\psi_j)dx$$

are mass matrix and stiffness matrix accordingly.

As in the differential case the following representations

$$[f,g]^N \overset{def}{=} (U^f, MU^g)(T)=\int_0^T[(IG^f,U_+^g)-(G_+^g,IU^f)](t)dt, \quad (14)$$

$$[f,g]_p^N \overset{def}{=} (U^f, KU^g)(T)=\int_0^T[(G^f,U_t^{g+})+(G_+^g,U_t^f)](t)dt. \quad (15)$$

holds. Emphasize that right hand sides of (14), (15) depend on $U^f|_{\Gamma_d\times 0,2T]}$ only, where $\Gamma_d$ is the set of boundary nodes.

Numerical experiments was made for unit disk $\Omega$ with continuous piecewise constant model of the density and the controls

$$f_{j\alpha}(x,t)=r(t-j\Delta t)q_\alpha(x),\ j=1,\ldots, N_t=T/\Delta t,\ \alpha=1,\ldots, N_b. \quad (16)$$

Here $r(t)$ is Ricker's impulse (fig.1), $q_\alpha$ are continuous piecewise "linear" functions on the boundary, $q_\alpha(x_\beta)=\delta_{\alpha\beta}, |x_\beta|=1, N_b$ is the number of boundary nodes. In what follows the basic controls (16) are numbered by one index $i=1,\ldots, N_b\times N_t$.

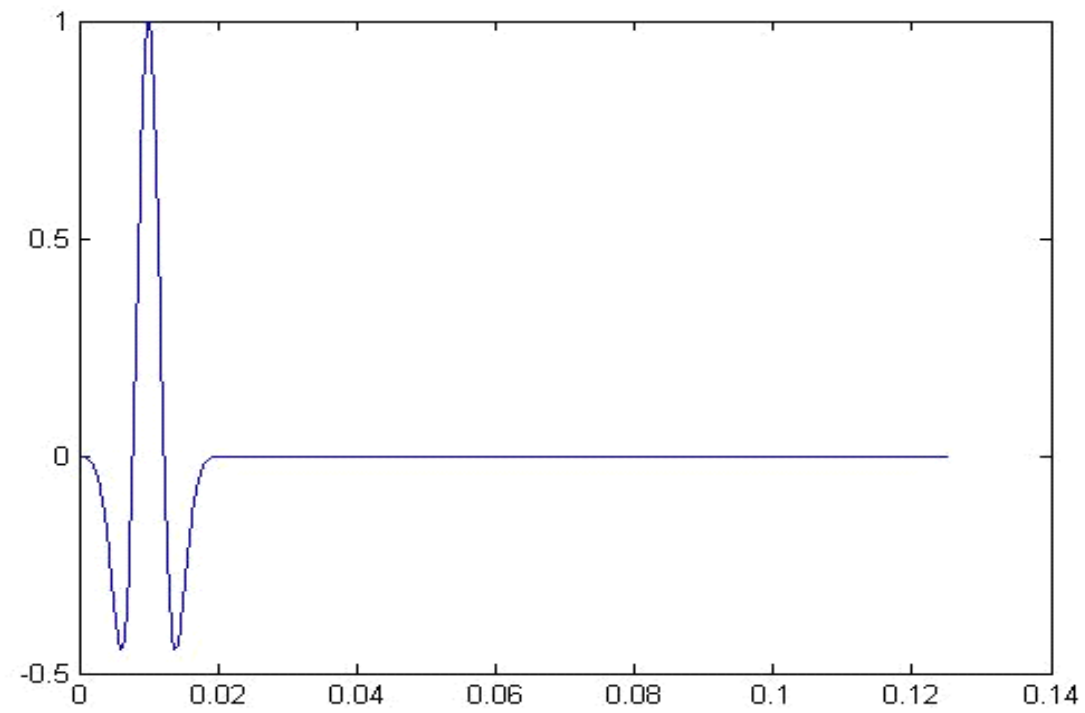

**1. Ricker wavelet**

The matrixes

$$C_{ij}=[f_i,f_j]^N,\ P_{ij}=[f_i,f_j]_p^N,\quad 1\le i,j\le N_b\times N_t$$

and values $U^{f_i}(T)|_{\Gamma_d}$ were used as the data of inverse problem. The reconstruction of $\rho$ was fulfilled using a scheme close to described above for differential inverse problem.

1. **Harmonic mesh functions.** Define harmonic mesh functions $\varphi_\alpha, \alpha=1,\ldots N_b$ as solutions of equations

$$K\varphi_\alpha=L_\alpha,\ \ \alpha=1,\ldots,\ N_b.$$

where linearly independent vectors $L_\alpha$ satisfy conditions of solvability

$$\sum_{x_i\in\Gamma_d}L_\alpha(x_i)=0,\alpha=1,\ldots,N_b.$$

The final harmonic mesh function is :

$$\varphi_{N_b+1}(x_j)=1,\ \ j=1,\ldots,N,\qquad K\varphi_{N_b+1}=0.$$

The $\varphi_\alpha$ is the mesh analogue of a harmonic function satisfying to the Neumann condition.

2. **The control problem.** Formally substituting $\varphi_\alpha$ for $U^f$, and $f_i$ for $g$ in (15) we get equations

$$[f_i,f]_p^N=(U^{f_i},L_\alpha)(T),i=1,\ldots,N_b\times N_t \tag{17}$$

w.r.t. $f$. To find a control, which gives $U^f(T)\approx\varphi_\alpha$ we have to use also equalities

$$U^f(x_i,T)=\varphi_\alpha(x_i),\ \ x_i\in\Gamma_d. \tag{18}$$

Denote by $f_\alpha$ the normal solution to the system (17),(18). The control $f_\alpha$ provided a well accuracy of equality $U^f(T)\approx\varphi_\alpha$ in each our numerical experiment (relative error in $l^2$ is about $10^{-8}$).

3. **Reconstruction**. Substituting $f$ for $f_\alpha,\varphi_\alpha$ for $U^f(T)$, $g$ for $f_\beta$, and $\varphi_\beta$ for $U^g$ in (14) we get

$$(\varphi_\alpha,M\varphi_\beta)=[f_\alpha,f_\beta]^N,\alpha,\beta=1,\ldots,N_{b+1}.$$

Let $\rho_k$ be the value of $\rho$ in the $k^{th}$ triangle, $k=1,\ldots,K$. Then we get the linear algebraic system w.r.t. $\rho_k$ :

$$\sum_{k=1}^{K}\rho_k\varphi_\alpha^i\varphi_\beta^j\int_{\Delta_k}\psi_i(x)\psi_j(x)dx=[f_\alpha,f_\beta]^N,\ \ \alpha,\beta=1,\ldots,\ N_b+1. \tag{19}$$

This system was sometimes ill-conditioned. Therefore we used natural a priori limitations for values $\rho_k$ and optimization algorithms to reconstruction. Below the results of numerical experiments are represented, $\delta$ means relative error in $l^2$.

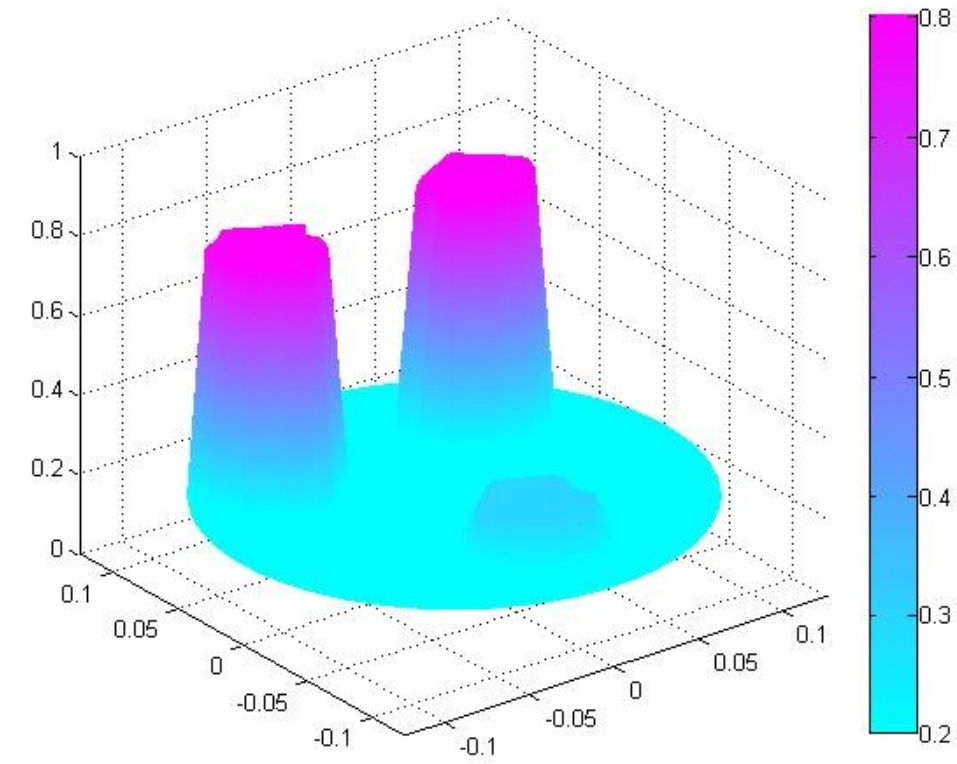

**2. Sample 1. Inclusions**

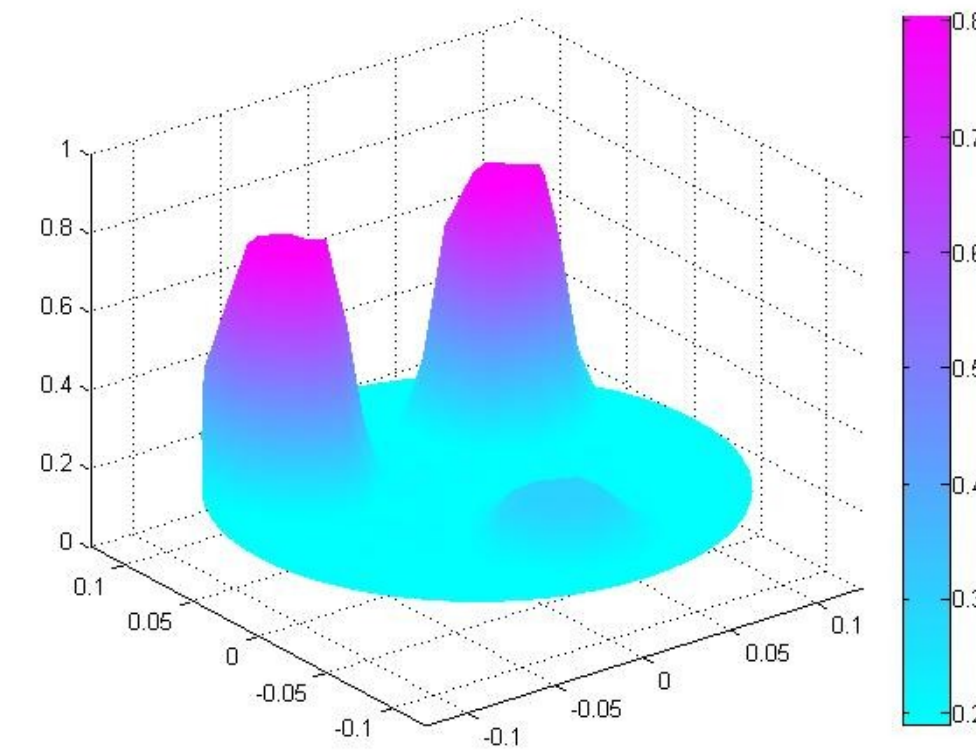

**3. Reconstruction of sample 1,** $\delta = 2,5\%$

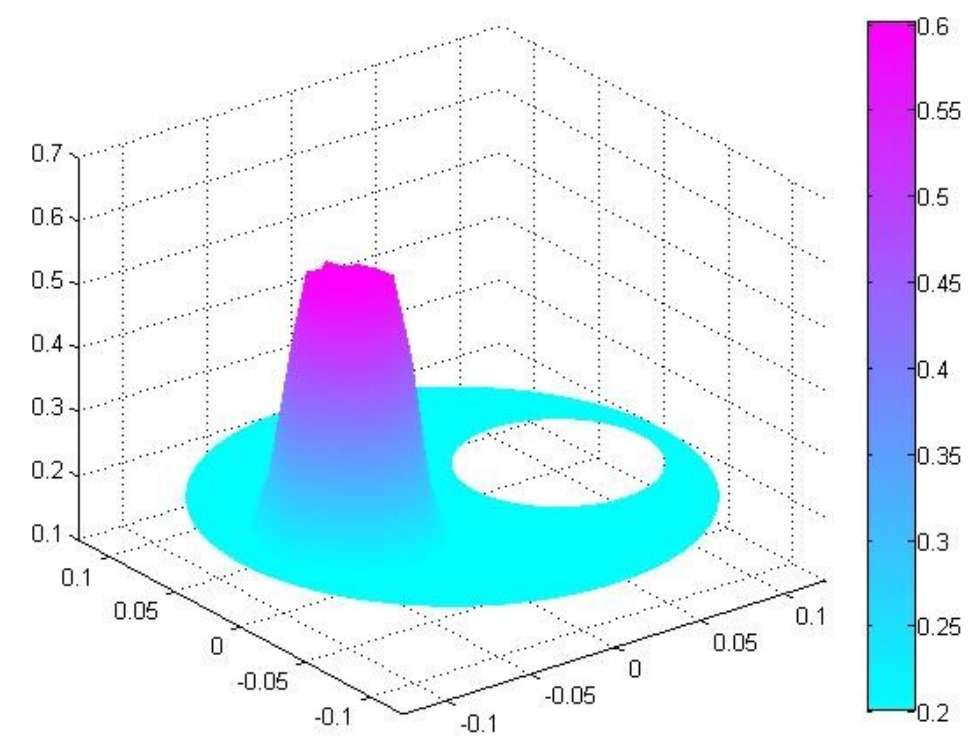

**4. Sample 2. Free inner boundary**

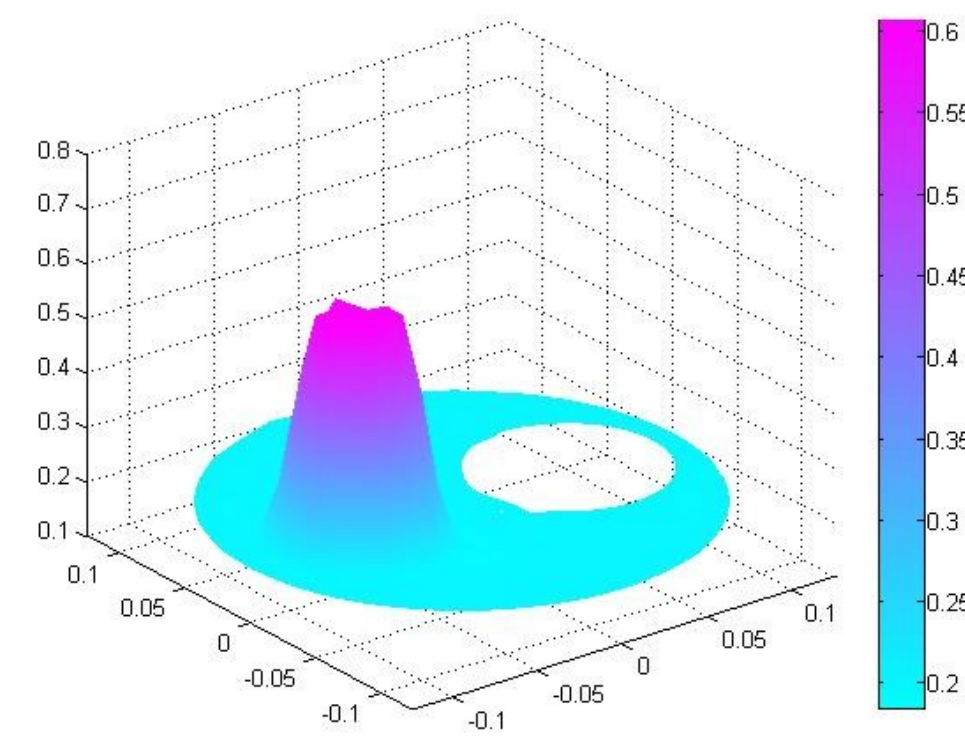

**5. Reconstruction of sample 2,** $\delta = 11\%$

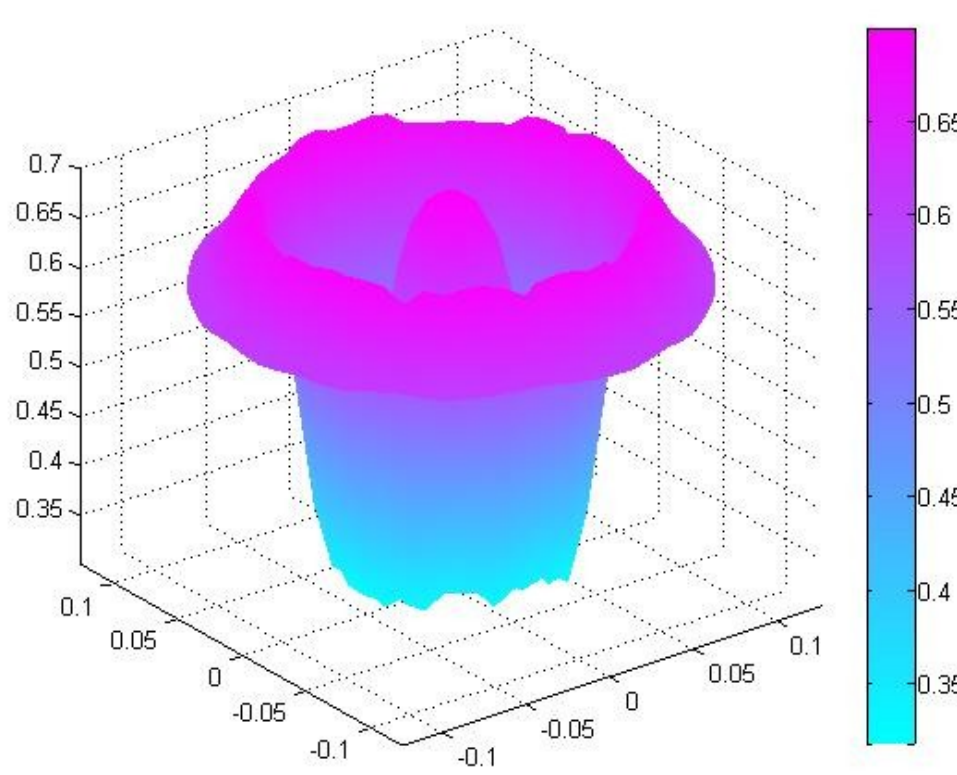

**6. Sample 3. Waveguides**

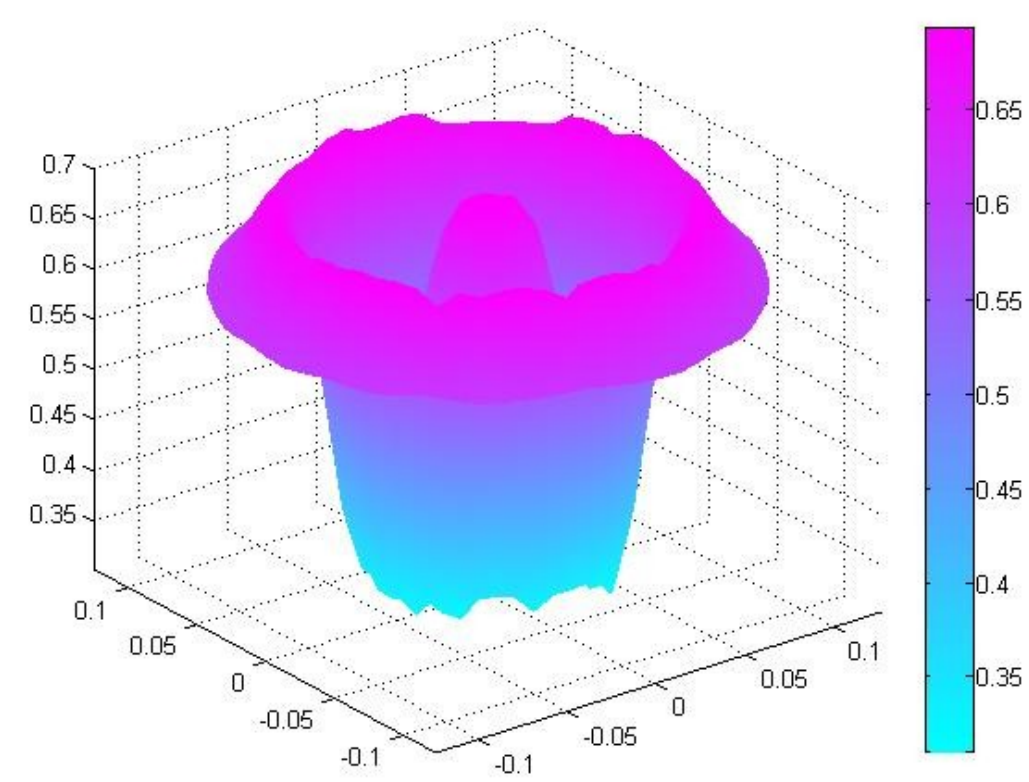

**7. Reconstruction of sample 3,** $\delta = 2,5\%$

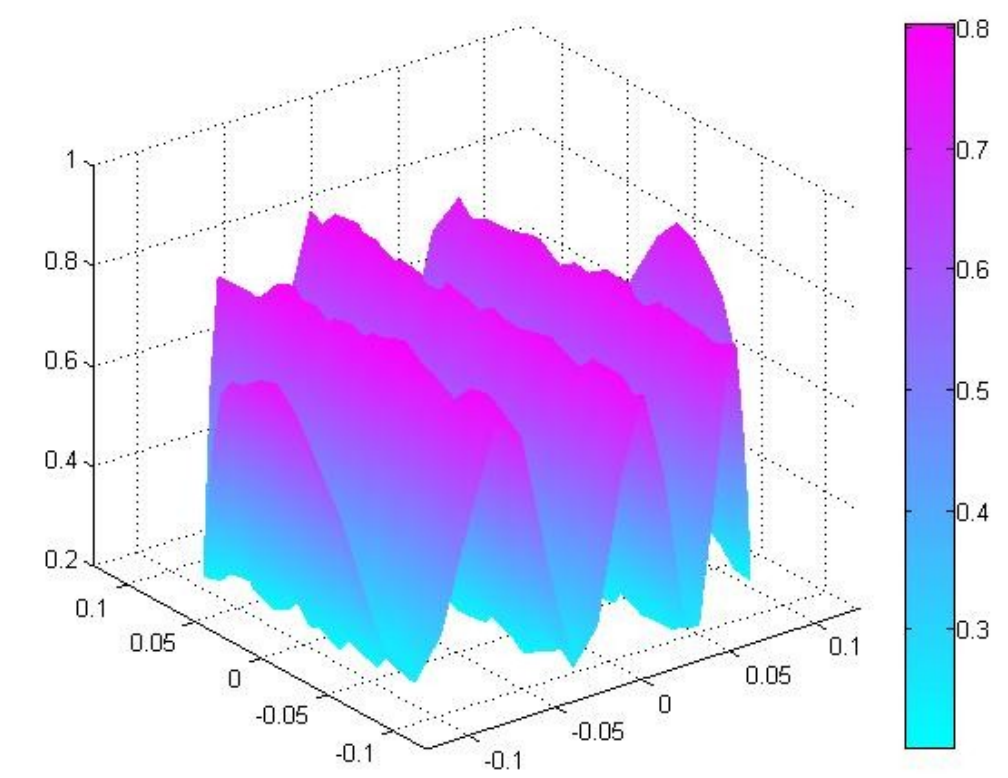

**8. Sample 4. Folds**

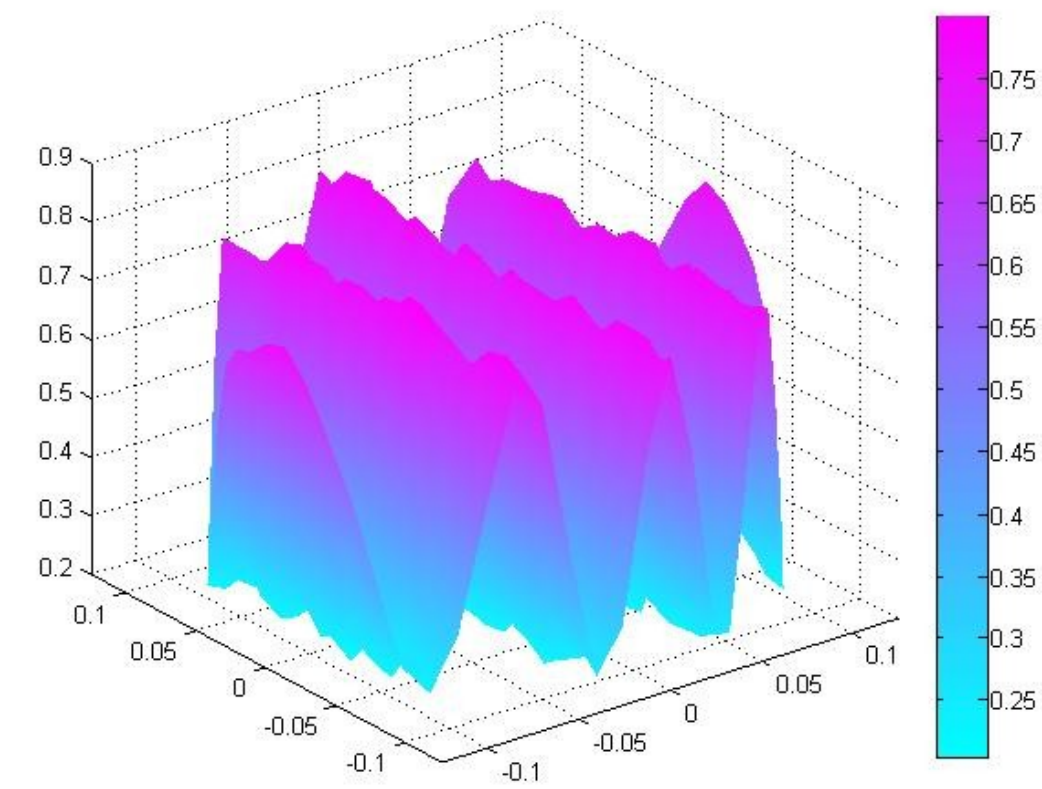

**9. Reconstruction of sample 4,** $\delta = 3\%$

## Acknowledgements

We would like to thank M. Belishev for the useful consultations and collaboration.